\documentstyle[11pt,newpasp,twoside,epsf]{article}
\def\edcomment#1{\iffalse\marginpar{\raggedright\sl#1\/}\else\relax\fi}
\reversemarginpar

\begin{document}
\title{Chemical abundances of $\beta$ Cephei stars from low- and high-resolution UV spectra}
\author{Jadwiga Daszy{\'n}ska-Daszkiewicz$^{1,2}$, Ewa Niemczura$^1$}
\affil{
$^1$Astronomical Institute of the Wroc{\l}aw University, ul. Kopernika 11, 51-622 Wroc{\l}aw, Poland\\
$^2$Instituut voor Sterrenkunde, Celestijnenlaan 200 B, B-3001 Leuven, Belgium}

\begin{abstract}
We determined stellar parameters ($T_{\rm eff}$, [m/H], $\theta$) and $E(B-V)$
for all $\beta$ Cephei stars observed during the IUE satellite mission.
All parameters were derived by using an algorithmic procedure of
fitting theoretical flux distribution (Kurucz 1996) to the IUE observations.
\end{abstract}

\section{Introduction}
In early type stars the ultraviolet spectral region is important for several reasons.
Firstly, since the majority of the total flux is emitted here,
it provides a rather sensitive indicator of a photospheric temperature and luminosity.
Secondly, the UV spectra are dominated by lines of the iron-group elements,
which cause the origin of the opacity bump at temperature $2 \times 10^5$~K, where the
classical $\kappa$ mechanism drives pulsations of $\beta$ Cep stars.
Consequently, the position of the instability domain in the HR diagram strongly depends
on the metal abundance (Pamyatnykh 1999).

We derive stellar parameters  (metallicity, [m/H], effective temperature, $T_{\rm eff}$,
stellar diameter, $\theta$) and interstellar extinction, $E(B-V)$, for all $\beta$ Cep stars
observed by {\it International Ultraviolet Explorer} satellite. The parameters are derived
by means of an algorithmic procedure of fitting theoretical flux distributions to the
low-resolution IUE spectra and optical spectrophotometric observations.
The errors are estimated by using the bootstrap method (Press et al. 1992).
We also show some examples of high-resolution HST/GHRS spectra for
one $\beta$~Cep star: $\gamma$ Peg.

\section{Analysis of low resolution IUE spectra}

\subsection{Observations and method}

The observational material consists of the IUE spectra with the absolute calibration and reduction done
by NEWSIPS package. The spectra were coadded if more than one spectrum was available for a star.
The ultraviolet observations expressed in the absolute units were supplemented
by optical spectrophotometric measurements taken from the literature. We used Johnson and Str\"omgren
photometry if no optical spectrophotometric data were available.

We applied the least-squares (LS) optimalization algorithm of fitting theoretical
flux distributions to observations. This method enables us to obtain various parameters
involved in stellar spectra ($T_{\rm eff}$, [m/H], $E(B-V)$, $\theta$). We used theoretical
models of Kurucz (1996) with the value of the microturbulent velocity, $v_t = 2$~km/s,
for all stars. In addition, the mean interstellar reddening curve of Fitzpatrick (1999)
was adopted for the majority of the analysed objects. Because of the spatial variability
of the extinction law, for the field stars with $E(B-V) > 0.10$, five additional parameters,
specifying the shape of the UV extinction curve (Fitzpatrick 1999) were estimated.
For $\beta$~Cep stars in open clusters we adopted the extinction curves from Massa \& Fitzpatrick (1986).
The surface gravity was determined as a mean value from four methods. Three of them use Str\"omgren
and Geneva photometry. In the last method, we estimated the gravities using the formula
obtained from stellar evolutionary models computed by A. Pamyatnykh (private communication)
for OPAL opacities with $Z = 0.02$.
We found the following relation: $\log g = -12.5894 + 4.4810\log T_{\rm eff} - 0.7870 \log L/L_{\odot}$,
with the standard deviation amounts to 0.01~dex. We used Hipparcos parallaxes in order to calculate
$\log L/L_\odot$. During the best-fit procedure the luminosity was corrected for the Lutz-Kelker bias
(Lutz \& Kelker 1973). The distances to the clusters were taken from the literature. The errors
were estimated by using the technique of bootstrap resampling (Press et al. 1992, Niemczura 2003).
This technique allows to get reliable uncertainties of the parameters and correlations between them.

\subsection{Results}

In Fig. 1, we show the $\log g$ vs. $\log T_{\rm eff}$ diagrams with observational points and theoretical
instability strips for three values of $Z$.
The mean value of the metal abundance parameter for the field stars  is equal
to $-0.14 \pm 0.03$~dex, and ranges from $-0.47$ (27 CMa) to $0.21$~dex (HN Aqr).
The mean values of [m/H] for the stars in clusters are equal to $0.05 \pm 0.06$~dex
for NGC~3293, $-0.43 \pm 0.05$ for NGC~4755 and $-0.01 \pm 0.06$ for NGC~6231.
The clusters NGC~3293 and NGC~4755 have similar ages,
but the $\beta$~Cep instability strip in the second one is shifted to the lower temperatures.
This can be explained by the lower value of [m/H] obtained for the NGC~4755 stars.
The mean [m/H] for all stars we analysed is  $-0.13 \pm 0.03$.
These determinations are consistent with our previous results (Daszy\'nska et al. 2002),
obtained with the less accurate method.  Metal abundances of hot stars in the solar
vicinity are lower by about 0.20~dex than the solar value and were reported by many authors
(see Niemczura 2003 and references therein). The values from the low-resolution IUE observations
are in agreement with these results. In the next step we checked whether the obtained metallicities
are correlated with any stellar parameter. We found small correlations between [m/H] and all parameters
(from 0.17 for $E(B-V)$ to 0.26 for $T_{\rm eff}$), which mean that [m/H] can be reliably derived
from the best-fit procedure. Then we tried to find quantities, which can be determined
by metallicity in $\beta$~Cep stars. The value of the dominant period, its amplitudes of
the light and radial velocity variations, as well as the projected rotational velocity
are not correlated with the value of [m/H], as has been already shown by Daszy\'nska et al. (2002).
For more details we refer reader to Niemczura \& Daszy\'nska-Daszkiewicz (2003).
\begin{figure}
\plotone{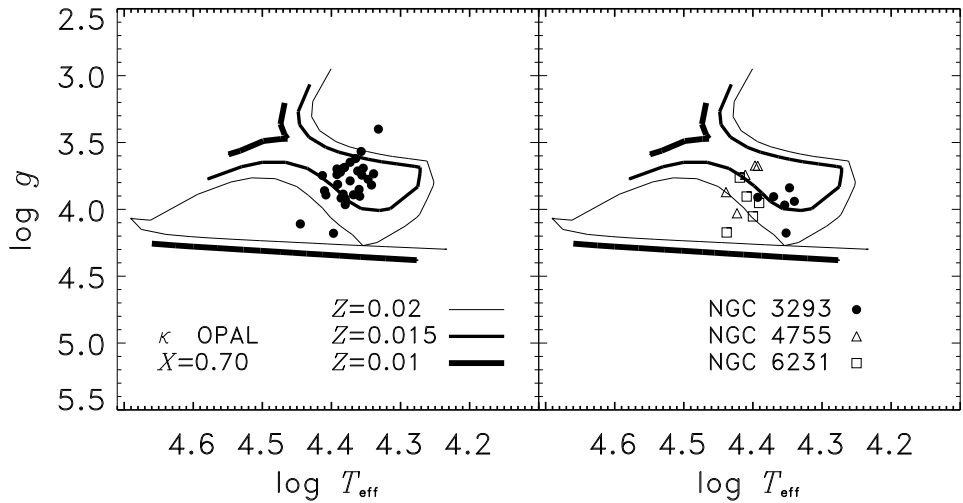}
\caption{Theoretical instability domains of $\beta$~Cep stars and observational data
in the $\log g$ -- $\log T_{\rm eff}$ diagram. The instability domains and lines of ZAMS were taken from
Pamyatnykh (1999).}
\end{figure}

\section{High resolution HST/GHRS spectra of the $\beta$~Cephei star $\gamma$~Peg}

High-quality UV spectra of $\gamma$~Peg have become available with Goddard High Resolution
Spectrograph (GHRS) aboard HST. Observations were made with two gratings, G160M (1340 -- 1380~{\AA})
and ECH-B (2060 -- 2071~{\AA}), with the spectral resolving power of about 20.000 and 80.000,
respectively. Both spectral ranges are dominated by the iron peak elements like
Fe, Cr, Mn, Ni, Si, Zn, V and Ti. In Fig. 2 we show a comparison of the ECH-B spectrum
of $\gamma$~Peg with the theoretical one computed with Kurucz LTE line-blanketed models,
for [m/H]$=-0.2$~dex and stellar parameters obtained from the above analysis of IUE spectra.
The overall fit is quite good, but for more detail analysis the abundances of elements
should be determined individually.
\begin{figure}
\plotone{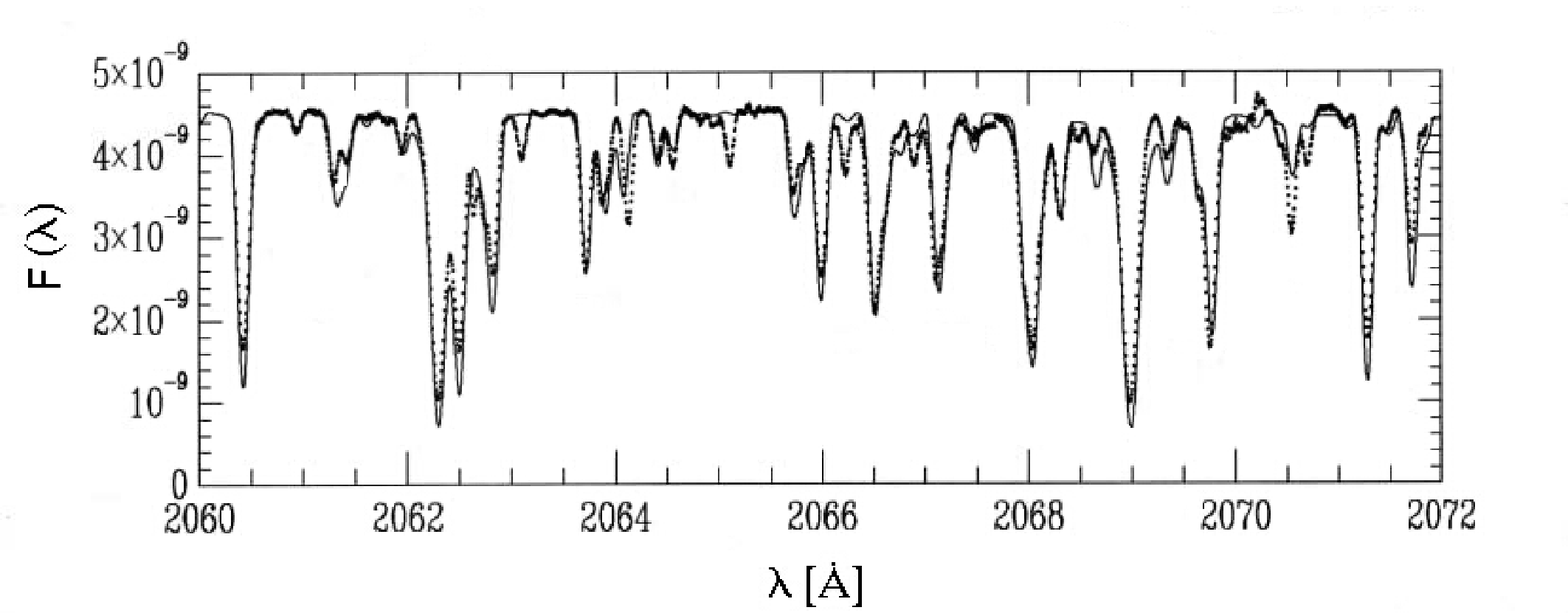}
\caption{Comparison of HST spectrum of $\gamma$ Peg with the theoretical one computed with
Kurucz LTE models for [m/H]$=-0.2$.}
\end{figure}

\section{Conclusions and future prospects}

The abundance of the metals, especially of iron, in $\beta$~Cephei pulsating variables is one
of the fundamental parameters. In the case of main sequence B stars, values obtained from the
UV spectra give mainly the estimation of [Fe/H]. Because the Kurucz models are overestimated in the
iron element we have to include the correction for [m/H] of about 0.12 dex. The metallicities of
the $\beta$~Cep stars predicted by the theory of pulsations are not in contradiction with most values determined
by us. Pamyatnykh (1999) showed that the $\beta$~Cephei instability strip vanishes for $Z \approx 0.01$,
corresponding to [m/H] $\approx -0.30$ dex, and to [m/H] $\approx -0.42$~dex from the Kurucz's fluxes.
There is a few stars with the metallicity parameter lower than this limit. But we have to remember that our values
give an information mainly about photospheric metal abundances.
Some other effects which we did not take into account, like diffusion or element mixing may also be importantant.
We did not change the mixture of elements, neither.
In spite of this, our results provide a very important information about the metallicity
range for these pulsating stars. However, for asteroseismological purposes, there is a need for a detailed
analysis of chemical composition, because oscillation frequencies are very sensitive to the adopted mixture.
This is the aim of our future work and here we show an example of high-resolution HST/GHRS spectrum
of $\gamma$~Peg with the best-fit of theoretical models.

\acknowledgments

We gratefully thank  Alosza Pamyatnykh  for  calculations a number of evolutionary  tracks.

\end{document}